\documentclass[conference]{IEEEtran}
\IEEEoverridecommandlockouts
\usepackage{cite}
\usepackage{amsmath,amssymb,amsfonts}
\usepackage{algorithmic}
\usepackage{algorithm}
\usepackage{graphicx}
\usepackage{textcomp}
\usepackage{xcolor}
\def\BibTeX{{\rm B\kern-.05em{\sc i\kern-.025em b}\kern-.08em
    T\kern-.1667em\lower.7ex\hbox{E}\kern-.125emX}}
\begin{document}

\title{Exploring the Feasibility of Automated Data Standardization using Large Language Models for Seamless Positioning\\
}

\author{
\IEEEauthorblockN{Max J. L. Lee}
\IEEEauthorblockA{
\textit{Department of Aeronautical and} \\
\textit{Aviation Engineering} \\
\textit{The Hong Kong Polytechnic University}\\
Hong Kong \\
maxjl.lee@connect.polyu.hk}
\and
\IEEEauthorblockN{Ju Lin}
\IEEEauthorblockA{
\textit{Department of Aeronautical and} \\
\textit{Aviation Engineering} \\
\textit{The Hong Kong Polytechnic University}\\
Hong Kong \\
ju.lin@connect.polyu.hk}
\and
\IEEEauthorblockN{Li-Ta Hsu}
\IEEEauthorblockA{
\textit{Department of Aeronautical and} \\
\textit{Aviation Engineering} \\
\textit{The Hong Kong Polytechnic University}\\
Hong Kong \\
lt.hsu@polyu.edu.hk}
}

\maketitle

\begin{abstract}
We propose a feasibility study for real-time automated data standardization leveraging Large Language Models (LLMs) to enhance seamless positioning systems in IoT environments. By integrating and standardizing heterogeneous sensor data from smartphones, IoT devices, and dedicated systems such as Ultra-Wideband (UWB), our study ensures data compatibility and improves positioning accuracy using the Extended Kalman Filter (EKF). The core components include the Intelligent Data Standardization Module (IDSM), which employs a fine-tuned LLM to convert varied sensor data into a standardized format, and the Transformation Rule Generation Module (TRGM), which automates the creation of transformation rules and scripts for ongoing data standardization. Evaluated in real-time environments, our study demonstrates adaptability and scalability, enhancing operational efficiency and accuracy in seamless navigation. This study underscores the potential of advanced LLMs in overcoming sensor data integration complexities, paving the way for more scalable and precise IoT navigation solutions.
\end{abstract}


\section{Introduction}
Accurate seamless positioning is paramount in the era of ubiquitous computing and the Internet of Things (IoT), enabling critical applications such as navigation, asset tracking, and location-based services \cite{ref1, ref2, ref3}. The proliferation of IoT devices has exponentially increased the demand for precise and reliable positioning systems. Researchers have explored various indoor positioning techniques, including Bluetooth Low Energy (BLE) beacons \cite{ref4}, Ultra-Wideband (UWB) ranging \cite{ref5}, and inertial sensor-based dead reckoning \cite{ref6}. Each method presents unique strengths and limitations regarding accuracy, coverage, scalability, and cost.

BLE beacons provide a cost-effective solution with reasonable accuracy but require dense deployment for high precision \cite{ref4}. UWB offers high accuracy and low latency, suitable for applications needing precise location information, such as industrial asset tracking \cite{ref5}, but is generally more expensive and has a limited range compared to BLE-based systems. Inertial sensor-based dead reckoning relies on accelerometers, gyroscopes, and magnetometers to estimate position changes but suffers from cumulative errors over time, necessitating periodic calibration with other positioning methods \cite{ref6}.

Fusing multiple positioning technologies has become a promising approach to enhance performance and robustness, especially in complex urban environments \cite{ref7}. Sensor fusion leverages the complementary strengths of different technologies to provide a more accurate and reliable positioning solution. For instance, combining BLE beacon data with inertial sensor data can compensate for the weaknesses of each method, resulting in improved accuracy and robustness.

However, integrating diverse positioning sensors poses significant challenges due to heterogeneous data formats and the need for sophisticated algorithms to handle uncertainties, noise, and interdependencies among different data sources \cite{ref9}. Traditional sensor fusion systems often rely on manual feature engineering and domain expertise, limiting their scalability and adaptability to new sensor types and environments \cite{ref11}.

Recent advancements in artificial intelligence (AI) and natural language processing (NLP), particularly Large Language Models (LLMs) like GPT-4-0613, offer promising solutions for the standardization of heterogeneous sensor data. These models have demonstrated remarkable capabilities in understanding and generating human-like text, inspiring researchers to explore their potential in other domains. Leveraging LLMs for automated data standardization can significantly reduce manual intervention and enhance scalability.

\subsection{Contributions}
This work makes several significant contributions to the field of seamless positioning and IoT applications:

\begin{itemize}
    \item \textbf{Innovative Application of LLMs}: Using LLMs for automating the standardization of heterogeneous sensor data is a new application extending their capabilities beyond traditional natural language processing tasks \cite{ref14}.
    \item \textbf{Enhanced Scalability and Adaptability}: Automating the data standardization process reduces the need for manual feature engineering and domain expertise, making the study more scalable and adaptable to new sensor types and environments.
    \item \textbf{Improved Accuracy}: Integrating standardized data with the Extended Kalman Filter (EKF) enhances the accuracy of the positioning system, providing more accurate and reliable positioning estimates.
\end{itemize}

\section{Feasibility Study Overview}
\begin{figure}[!t]
\centering
\includegraphics[width=3.5in]{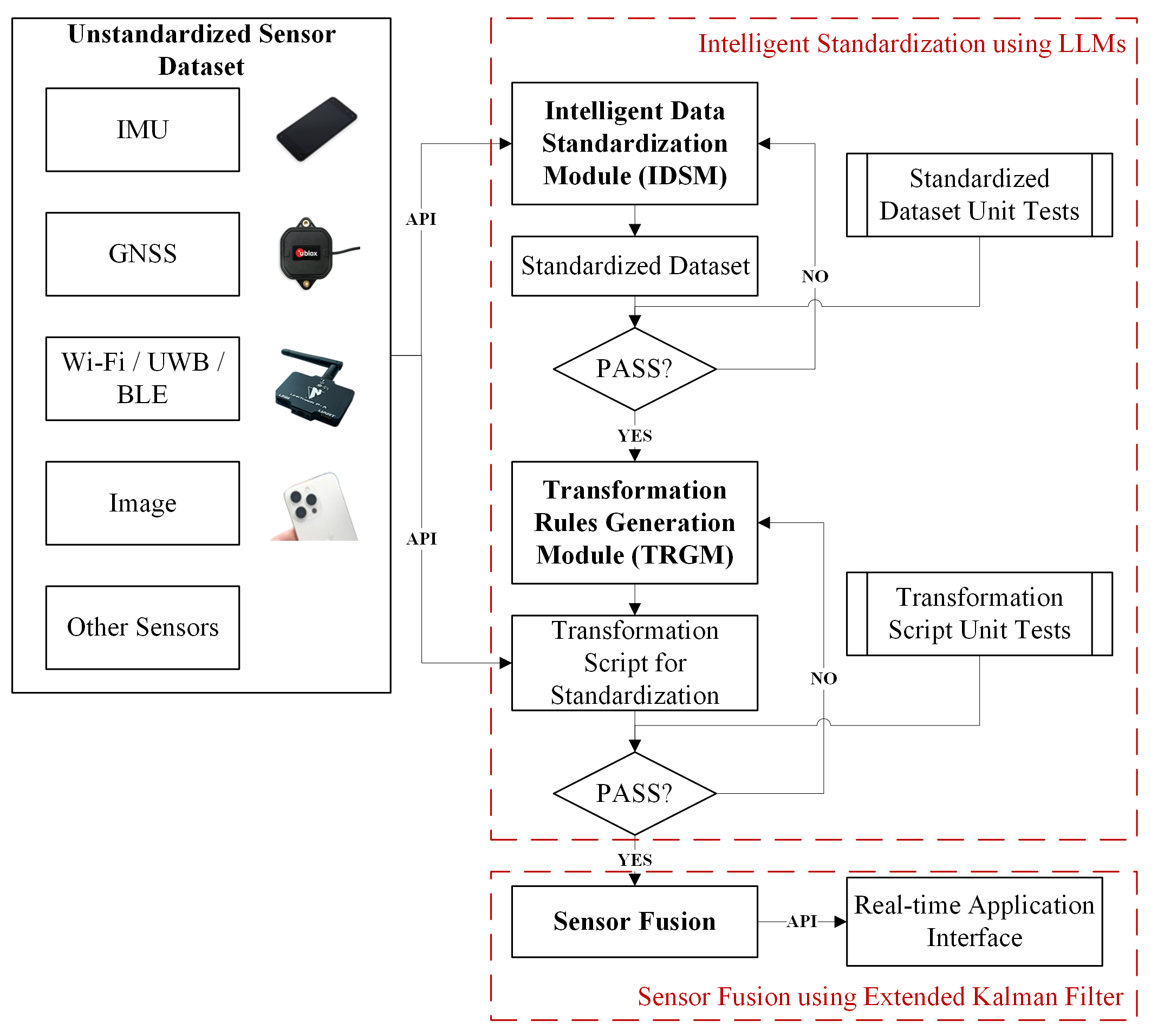}
\caption{Overview of the Proposed Feasibility Study for Automated Data Standardization and Sensor Fusion.}
\label{fig:flowchart}
\end{figure}

The flowchart of the proposed feasibility study, depicted in Fig. \ref{fig:flowchart}, illustrates the iterative standardization and validation process essential for enhancing seamless positioning systems.

The study begins with collecting unstandardized sensor data from sources like smartphones, IoT devices, and UWB tags. This heterogeneous data is processed by the Intelligent Data Standardization Module (IDSM), leveraging a fine-tuned Large Language Model (LLM) to automate standardization. The IDSM segments incoming data by sensor type and normalizes complex elements (e.g., timestamps to UNIX nanoseconds). The standardized data is formatted according to predefined specifications, as shown in Table \ref{tab:table1}.

Unit tests ensure the accuracy and integrity of the standardized data before progressing. Following standardization, the Transformation Rule Generation Module (TRGM) creates transformation rules, generating scripts for new sensor data standardization, enhancing scalability. The scripts undergo unit testing for functionality and reliability.

Finally, the Extended Kalman Filter (EKF) integrates standardized data from multiple sensors, improving the system's precision and robustness. Covariance matrices represent uncertainties, which are manually specified but may require adaptive approaches for real-world scenarios.

\begin{table}[!t]
\caption{Standardized Sensor Data Schema Structure}
\label{tab:table1}
\centering
\begin{tabular}{|p{1cm}|p{1cm}|p{2cm}|p{3cm}|}
\hline
\textbf{Sensor Type} & \textbf{Required Fields} & \textbf{Values Object Properties} & \textbf{Description} \\
\hline
Magneto-meter & name, time, values & x (µT), y (µT), z (µT) & Measures magnetic field strength along x, y, and z axes in microteslas (µT). \\
\hline
Gyro-scope & name, time, values & x (rad/s), y (rad/s), z (rad/s) & Measures angular velocity along x, y, and z axes in radians per second (rad/s). \\
\hline
Accelero-meter & name, time, values & x (m/s²), y (m/s²), z (m/s²) & Measures acceleration along x, y, and z axes in meters per second squared (m/s²). \\
\hline
Gravity & name, time, values & x (m/s²), y (m/s²), z (m/s²) & Measures gravity effects along x, y, and z axes in meters per second squared (m/s²). \\
\hline
Ultra-Wideband (UWB) & name, time, values & position (m) & Determines spatial position in meters (m). \\
\hline
Bluetooth & name, time, values & position (m) & Determines spatial position in meters (m). \\
\hline
Pedometer & name, time, steps & steps (count) & Tracks the number of steps taken (count). \\
\hline
Ori-entation & name, time, values & qx, qy, qz, qw & Provides orientation details in quaternion format. \\
\hline
Baro-meter & name, time, values & relative altitude (m), pressure (mBar) & Measures relative altitude in meters (m) and atmospheric pressure in millibars (mBar). \\
\hline
Location & name, time, values & latitude (°), longitude (°), altitude (m), speed (m/s), speed accuracy (m/s), horizontal accuracy (m), vertical accuracy (m) & Provides comprehensive location data including coordinates (degrees), speed (meters per second), altitude (meters), and accuracies (meters). \\
\hline
Image & name, time, image & image (data) & Provides image data in binary format. \\
\hline
\end{tabular}
\end{table}

\section{Intelligent Data Standardization Module (IDSM)}
The IDSM's primary objective is to transform heterogeneous sensor data \(\mathcal{D}\) into a standardized format \(\mathcal{S}\). The raw data collected from various sensors is represented as:

\begin{equation}
\mathcal{D} = \{d_1, d_2, \ldots, d_n\}
\label{eq:rawdata}
\end{equation}

where \(d_i\) denotes the data from the \(i\)-th sensor. The standardization process is expressed as:

\begin{equation}
\mathcal{S} = \mathcal{F}_{\text{IDSM}}(\mathcal{D})
\label{eq:standardization}
\end{equation}

The IDSM was fine-tuned using a curated dataset with 100 complex examples, addressing specific challenges in sensor data standardization. The standardized data schema structure is detailed in Table \ref{tab:table1}. Table \ref{tab:data_issues} highlights various edge cases and scenarios.

\begin{table}[!t]
\caption{Common Data Input Issues and Descriptions}
\label{tab:data_issues}
\centering
\begin{tabular}{|p{3cm}|p{5cm}|}
\hline
\textbf{Issue Type} & \textbf{Description} \\
\hline
Missing Values & Data entries were absent, requiring prediction or marking. \\
\hline
Irregular Data Formats & Sensor data appeared in non-uniform formats needing standardization. \\
\hline
Non-standard Unit Representations & Units of measurement varied, necessitating normalization. \\
\hline
Incomplete Information & Datasets were partially filled, reflecting real-world scenarios. \\
\hline
\end{tabular}
\end{table}

The IDSM was developed and fine-tuned using the Azure platform and the GPT-4-0613 model. Azure's robust infrastructure provided the computational resources necessary for handling large datasets and performing iterative validations efficiently. This integration allowed for seamless scaling and deployment of the IDSM, ensuring it could handle real-time data processing requirements.

\subsection{Training and Validation Performance} 
\begin{figure}[!t]
\centering
\includegraphics[width=3.5in]{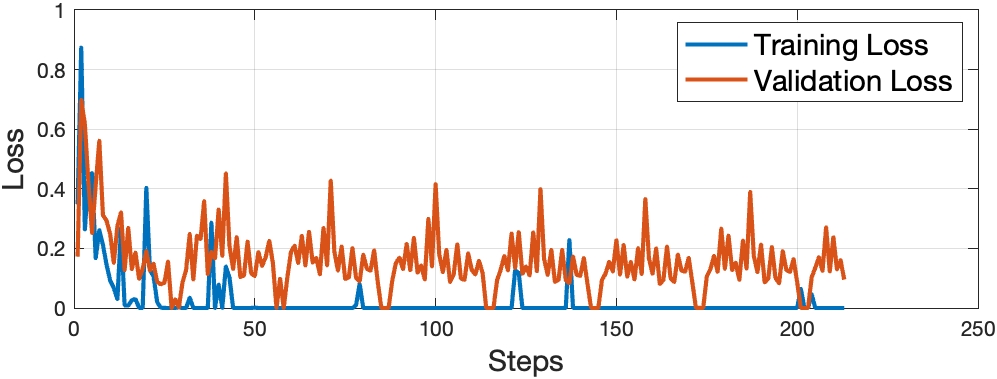}
\caption{Training and Validation Loss of the Intelligent Data Standardization Module (IDSM) over Steps.}
\label{fig:training_performance}
\end{figure}

The IDSM's performance was assessed through training and validation metrics. Initially, the training loss was 0.3484 (Fig.~\ref{fig:training_performance}), decreasing steadily to zero by step 27. Training accuracy started at 93.38\%, reaching 100\% by step 14 and maintaining this level.

Validation metrics showed similar trends. The validation loss started at 0.1724, briefly increased to 0.6984 at step 2, then declined to nearly zero by step 27. Validation accuracy began at 96.14\%, achieving near 100\% by step 27.

These results highlight the model's robust learning and generalization capabilities, effectively standardizing diverse sensor data in real-world scenarios.

\begin{figure}[!t]
\centering
\includegraphics[width=3.5in]{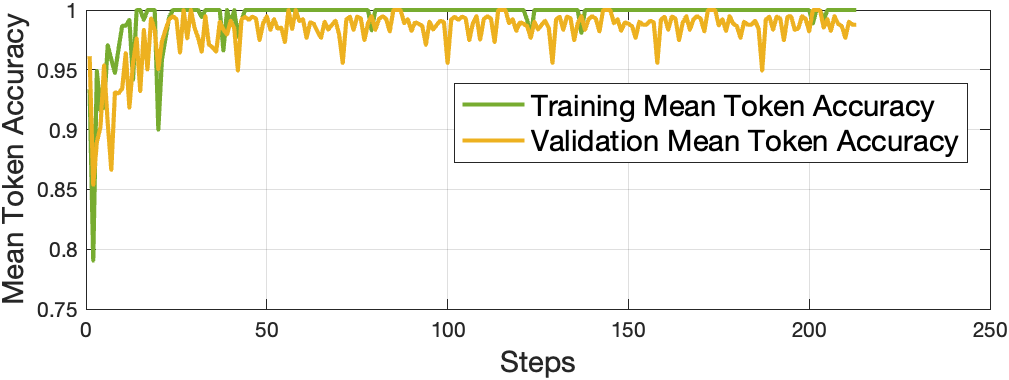}
\caption{Mean Token Accuracy of the Intelligent Data Standardization Module (IDSM) over Steps.}
\label{fig:meantokenaccuracy}
\end{figure}

\section{Standardized Dataset Unit Tests}
The unit tests validate the standardized dataset \(\mathcal{S}\) against predefined JSON schemas for various sensor types. The validation function is expressed as:

\begin{equation}
(\nu, e) = \mathcal{V}_{\text{IDSM}}(\mathcal{S}, \sigma)
\label{eq:validation}
\end{equation}

where \(\nu\) indicates whether \(\mathcal{S}\) conforms to schema \(\sigma\), and \(e\) contains details of validation errors. The iterative process ensures \(\mathcal{S}\) conforms to the schema through multiple cycles if necessary, up to a maximum of five iterations, based on empirical evidence and practical considerations.

Most datasets (24 out of 30) required only one iteration for successful validation. Two datasets required up to four iterations, while four datasets did not achieve successful validation within the iteration limit of five. This rapid initial convergence supports five iterations as an optimal limit.

These results underscore the robustness and effectiveness of the IDSM in standardizing diverse sensor data in most cases, ensuring high data quality and consistency. However, the datasets that did not validate within the iteration limit highlight areas for potential improvement in handling more complex data entries.

\section{Transformation Rules Generation Module (TRGM)}
The TRGM automates deriving transformation rules as detailed in Table \ref{tab:table2} for JSON structures using the GPT-4-0613 model. It converts input JSON files into a specified output format, reducing manual intervention. The process is represented as:

\begin{equation}
\mathcal{R} = \mathcal{F}_{\text{TRGM}}(\mathcal{S}, \mathcal{I})
\label{eq:trgm}
\end{equation}

where \(\mathcal{S}\) is the standardized data and \(\mathcal{I}\) is the input JSON structure. The transformation rules are then used to create a "Transformation Script for Standardization" \(\mathcal{T}\), ensuring consistency in data transformation tasks.

Table \ref{tab:table2} shows the schema structure for the transformation rules used by the TRGM.

\begin{table}[!t]
\caption{Transformation Rules Schema Structure}
\label{tab:table2}
\centering
\begin{tabular}{|p{1.5cm}|p{1.5cm}|p{4.5cm}|}
\hline
\textbf{Field Name} & \textbf{Data Type} & \textbf{Description} \\
\hline
\texttt{inputPath} & String & JSONPath expression pointing to the source field in the input JSON structure. \\
\hline
\texttt{outputPath} & String & JSONPath expression pointing to the target field in the output JSON structure. \\
\hline
\multicolumn{3}{|l|}{\textbf{Example}} \\
\hline
\texttt{inputPath} & String & \texttt{\$.sensor\_data.Accelerometer. timestamp} \\
\hline
\texttt{outputPath} & String & \texttt{\$[?(@.name == 'Accelerometer')].time} \\
\hline
\end{tabular}
\end{table}

\section{Transformation Script Unit Tests}
The unit tests for the TRGM validate the accuracy of the transformation scripts. The validation function is described as:

\begin{equation}
(\nu, e) = \mathcal{V}_{\text{TRGM}}(\mathcal{T}, \mathcal{S})
\label{eq:trgm_validation}
\end{equation}

where \(\nu\) indicates whether the output from \(\mathcal{T}\) matches the standardized data \(\mathcal{S}\), and \(e\) contains error details. The iterative validation process refines the transformation rules through successive refinements until the correct output is achieved.

\section{Sensor Fusion}
The Sensor Fusion module employs an Extended Kalman Filter (EKF) to integrate data from multiple sensors, providing accurate, real-time estimates of 3D positions and velocities \cite{ref16}.

\subsection{State Vector and Transition Matrix}
The state vector \(\mathbf{x}\) includes 3D positions and velocities:

\begin{equation}
\mathbf{x} = \begin{bmatrix} x \\ y \\ z \\ v_x \\ v_y \\ v_z \end{bmatrix}
\label{eq:state_vector}
\end{equation}

The state transition matrix \(\mathbf{F}_k\) governs the evolution of this state vector:

\begin{equation}
\mathbf{F}_k = \begin{bmatrix} 
\mathbf{I}_{3 \times 3} & \Delta t \mathbf{I}_{3 \times 3} \\ 
\mathbf{0}_{3 \times 3} & \mathbf{I}_{3 \times 3} 
\end
{bmatrix}
\label{eq:state_transition}
\end{equation}

where \(\Delta t\) is the time interval between updates.

\subsection{Measurement Model}
The EKF utilizes measurements from multiple sensors. Table \ref{table:measurements} summarizes the measurement vectors and covariance matrices.

\begin{table}[!t]
\caption{Measurement Vectors and Covariance Matrices for Sensors}
\label{table:measurements}
\centering
\begin{tabular}{|p{1.5cm}|p{2.5cm}|p{3.5cm}|}
\hline
\textbf{Sensor Type} & \textbf{Measurement Vector} & \textbf{Measurement Covariance Matrix} \\
\hline
GNSS Receiver & \( \mathbf{z}_{\text{GNSS}} = \begin{bmatrix} \phi_{\text{GNSS}} \\ \lambda_{\text{GNSS}} \\ h_{\text{GNSS}} \end{bmatrix} \) & 
\( \mathbf{R}_{\text{GNSS}} = \begin{bmatrix} 
655.00 & 0 & 0 \\ 
0 & 655.00 & 0 \\ 
0 & 0 & 655.00 
\end{bmatrix} \) \\
\hline
UWB Sensor & \( \mathbf{z}_{\text{UWB}} = \begin{bmatrix} x_{\text{UWB}} \\ y_{\text{UWB}} \\ z_{\text{UWB}} \end{bmatrix} \) & 
\( \mathbf{R}_{\text{UWB}} = \begin{bmatrix} 
1.00 & 0 & 0 \\ 
0 & 1.00 & 0 \\ 
0 & 0 & 1.00 
\end{bmatrix} \) \\
\hline
Camera & \( \mathbf{z}_{\text{cam}} = img_{\text{cam}} \) & 
\( \mathbf{R}_{\text{cam}} = \begin{bmatrix} 
0.15 & 0 & 0 \\ 
0 & 0.15 & 0 \\ 
0 & 0 & 0.15 
\end{bmatrix} \) \\
\hline
\end{tabular}
\end{table}

\subsubsection{Measurement Variance Calculation}
Measurement variances are derived from experimental results. For the GNSS receiver:

\begin{equation}
\sigma_{\text{GNSS}}^2 = 25.02^2 + 5.47^2 = 655.00 \, \text{m}^2
\label{eq:gnss_variance}
\end{equation}

For the UWB sensor:

\begin{equation}
\sigma_{\text{UWB}}^2 = 0.79^2 + 0.62^2 = 1.00 \, \text{m}^2
\label{eq:uwb_variance}
\end{equation}

For the camera:

\begin{equation}
\sigma_{\text{cam}}^2 = 0.32^2 + 0.23^2 = 0.15 \, \text{m}^2
\label{eq:cam_variance}
\end{equation}

\subsection{Control Input}
The control input vector \(\mathbf{u}\) includes accelerations, angular velocities, and magnetometer readings from the IMU. The orientation matrix \(\mathbf{C}\) transforms the accelerations from the IMU's frame to the NED frame. The control input matrix \(\mathbf{B}_{\text{IMU}}\) integrates these transformed accelerations into the state vector, accounting for the time step \(\Delta t\).

\begin{table}[!t]
\caption{Control Input Vectors and Process Noise Covariance for IMU}
\label{table:control}
\centering
\begin{tabular}{|p{0.7cm}|p{1.3cm}|p{1.5cm}|p{3.5cm}|}
\hline
\textbf{Sensor Type} & \textbf{Control Input Vector} & \textbf{Control Input Matrix} & \textbf{Process Noise Covariance} \\
\hline
IMU & \( \mathbf{u}_{\text{IMU}} = \begin{bmatrix} a_x \\ a_y \\ a_z \\ \omega_x \\ \omega_y \\ \omega_z \\ m_x \\ m_y \\ m_z \end{bmatrix} \) & 
\( \mathbf{B}_{\text{IMU}} = \begin{bmatrix} 
\frac{\Delta t^2}{2} \mathbf{I}_{3 \times 3} \\ 
\Delta t \mathbf{I}_{3 \times 3} 
\end{bmatrix} \) & 
\( \mathbf{Q} = \begin{bmatrix} 
\frac{\sigma_a^2 \Delta t^4}{4} \mathbf{I}_{3 \times 3} & \frac{\sigma_a^2 \Delta t^3}{2} \mathbf{I}_{3 \times 3} \\ 
\frac{\sigma_a^2 \Delta t^3}{2} \mathbf{I}_{3 \times 3} & \sigma_a^2 \Delta t^2 \mathbf{I}_{3 \times 3} 
\end{bmatrix} \) \\
\hline
\end{tabular}
\end{table}

\section{Experiment}
\subsection{Experiment Setup}
\begin{figure}[!t]
\centering
\includegraphics[width=3.5in]{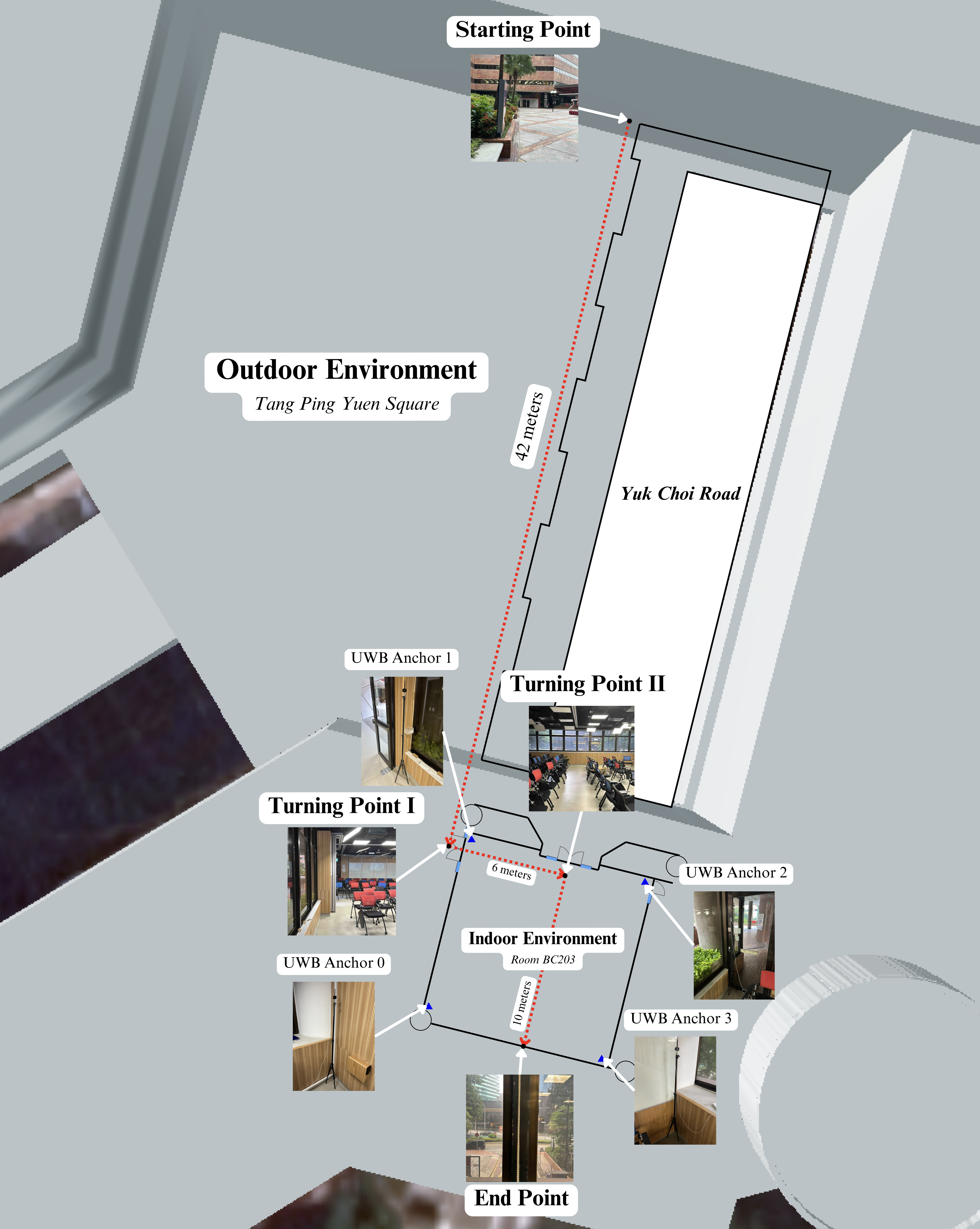}
\caption{Experiment Path and Ultra-Wideband (UWB) Setup for Data Collection.}
\label{fig:experiment_path}
\end{figure}

The experiment was conducted in a dynamic environment transitioning from outdoors to indoors, as depicted in Fig.~\ref{fig:experiment_path}. This location, characterized by tall buildings, represents urbanized areas where enhanced positioning accuracy through sensor fusion is essential. The total length of the ground truth path was approximately 60 meters. Data collection was performed using four different sensors, each on dedicated devices to simulate sensor fusion from various sources. Table~\ref{table:sensors} lists the sensors and their fused counterparts.

\begin{table}[!t]
\caption{Descriptions of Measurement Sensors and Fusion Methods}
\label{table:sensors}
\centering
\begin{tabular}{|p{1.5cm}|p{6.5cm}|}
\hline
\textbf{Method} & \textbf{Description} \\
\hline
Ground Truth (GT) & Data collected at landmark locations identified on Hong Kong Lands Department 3D Spatial Data 3D-BIT00, with an accuracy within 1-2 meters, combined with the indoor floor plan from The Hong Kong Polytechnic University. \\
\hline
GNSS \cite{ref16} & The u-blox F9P receiver connected to a computer via serial port, streaming GNSS location data at one position per second. A custom script parsed NMEA sentences and transmitted the data in real-time to an API endpoint. \\
\hline
UWB \cite{ref17} & Four Nooploop LinkTrack P-A Series UWB anchors were placed at the corners of the indoor environment, positioned 1.5 meters above the ground. The system was calibrated to align with the global coordinates of the floor plan, enabling real-time global positioning. Data from the UWB system was transmitted at 20 Hz using the Robot Operating System (ROS), which broadcast the data to an API endpoint. \\
\hline
VPS \cite{ref18,ref19} & A 3D sparse point cloud was constructed for positioning using the hierarchical localization toolbox (hloc), consisting of key points covering the area and registered to the global map. A Samsung Galaxy Note 20 Ultra smartphone captured images at 1 frame per second in real-time and sent them to an API endpoint, where the server determined the device's position relative to the geo-registered point cloud. \\
\hline
IMU \cite{ref20} & The iPhone 14 Pro used the “Sensor Logger” app to stream IMU data at 100 Hz to the API endpoint, including accelerometer, gyroscope, and magnetometer readings to estimate the device's orientation and motion. \\
\hline
GNSS + IMU & A GNSS solution using the u-blox F9P receiver fused with iPhone IMU data. \\
\hline
VPS + IMU & VPS positioning method fused with iPhone IMU data. \\
\hline
UWB + IMU & A UWB-based positioning method fused with iPhone IMU data. \\
\hline
GNSS + VPS + UWB + IMU & A comprehensive solution combining GNSS (u-blox F9P), VPS, UWB, and iPhone IMU data. \\
\hline
\end{tabular}
\end{table}

\subsection{Experiment Results}
The proposed feasibility study was rigorously tested using sensor data collected from multiple sources, including smartphones, IoT devices, and UWB tags. The initial dataset comprised 10,000 unstandardized data entries, each containing various elements such as timestamps, sensor readings, and device identifiers.

\subsubsection{Positioning System Accuracy}
\begin{figure}[!t]
\centering
\includegraphics[width=3.5in]{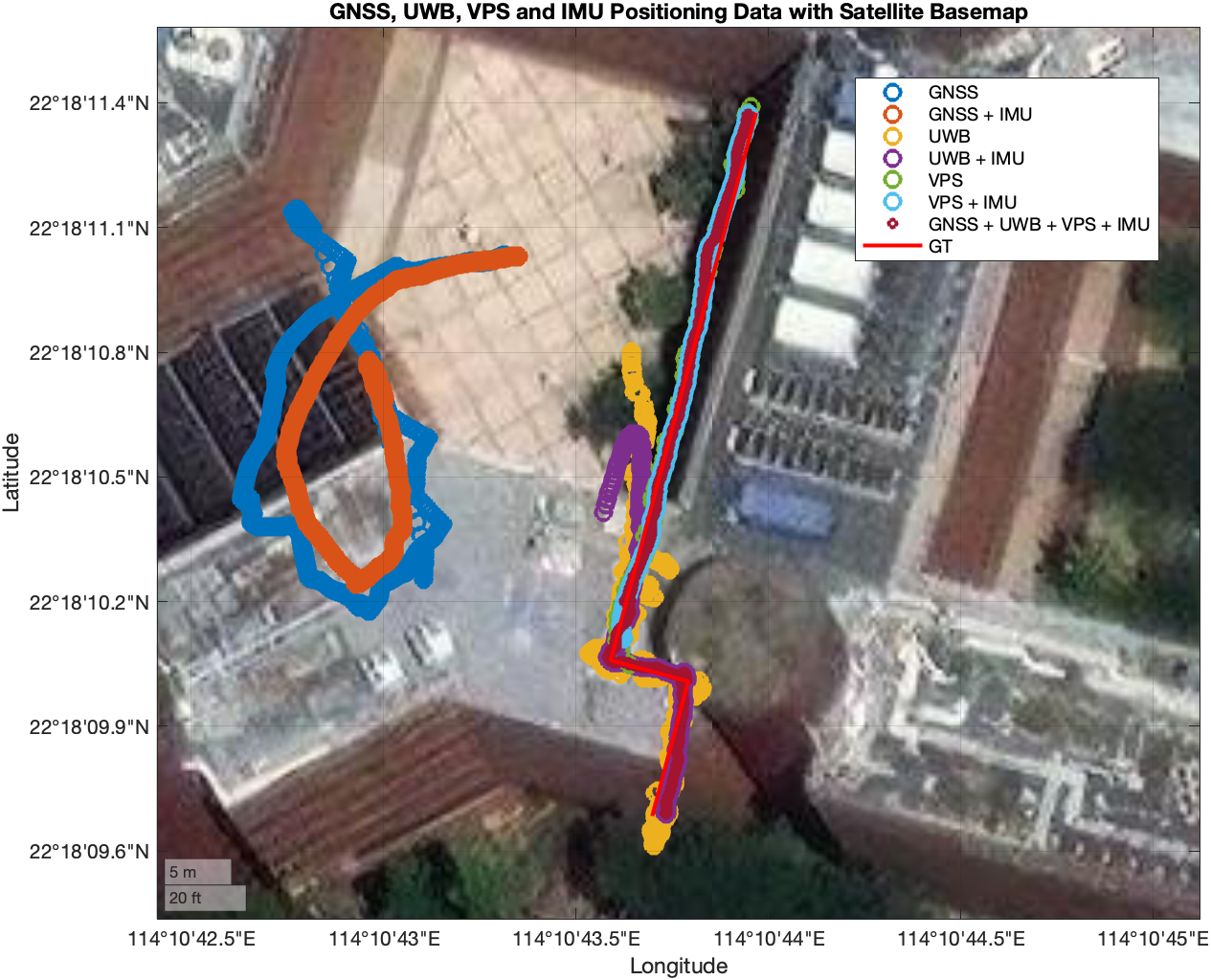}
\caption{Comparison of Positioning Results from Various Methods with Ground Truth.}
\label{fig:positioning_results}
\end{figure}

Positioning accuracy was evaluated by comparing the results obtained from different methods against the ground truth data. The error was measured by calculating the shortest distance between our solution and the ground truth path. The results are presented in Table \ref{tab:measurement_results} and visualized in Fig.~\ref{fig:positioning_results}.

\begin{table}[!t]
\caption{Performance Metrics of Different Positioning Methods}
\label{tab:measurement_results}
\centering
\begin{tabular}{|p{1cm}|p{1cm}|p{1cm}|p{1cm}|p{1cm}|p{1cm}|}
\hline
\textbf{Method} & \textbf{Mean Error (m)} & \textbf{Std. Dev. (m)} & \textbf{RMSE (m)} & \textbf{Median Error (m)} & \textbf{Max Error (m)} \\
\hline
GNSS & 25.02 & 5.47 & 25.61 & 25.21 & 33.75 \\
\hline
GNSS + IMU & 23.24 & 4.22 & 23.62 & 23.38 & 29.40 \\
\hline
UWB & 0.79 & 0.62 & 1.00 & 0.71 & 4.75 \\
\hline
UWB + IMU & 0.69 & 0.89 & 1.13 & 0.31 & 3.92 \\
\hline
VPS & 0.32 & 0.23 & 0.39 & 0.30 & 0.76 \\
\hline
VPS + IMU & 0.33 & 0.23 & 0.41 & 0.31 & 0.91 \\
\hline
GNSS + VPS + UWB + IMU & 0.33 & 0.24 & 0.41 & 0.27 & 0.95 \\
\hline
\end{tabular}
\end{table}

Table \ref{tab:measurement_results} and Fig. \ref{fig:positioning_results} demonstrate the benefits of integrating multiple positioning technologies. The standalone GNSS system shows a mean error of 25.02 meters, revealing significant deviations due to urban multipath effects. Integrating GNSS with an IMU reduces the mean error to 23.24 meters, but errors remain substantial. The UWB system shows improved performance indoors, with a mean error of 0.79 meters, further reduced to 0.69 meters when combined with an IMU. The Visual Positioning System (VPS) shows exceptional accuracy in outdoor settings, with a mean error of 0.32 meters. When integrated with an IMU, VPS maintains high accuracy with a mean error of 0.33 meters. Combining GNSS, VPS, UWB, and IMU leverages the strengths of each technology, providing reliable and accurate real-time positioning in urban settings. The GNSS + VPS + UWB + IMU combination achieves a mean error of 0.33 meters, demonstrating the effectiveness of this integrated approach.

\begin{figure}[!t]
\centering
\includegraphics[width=3.5in]{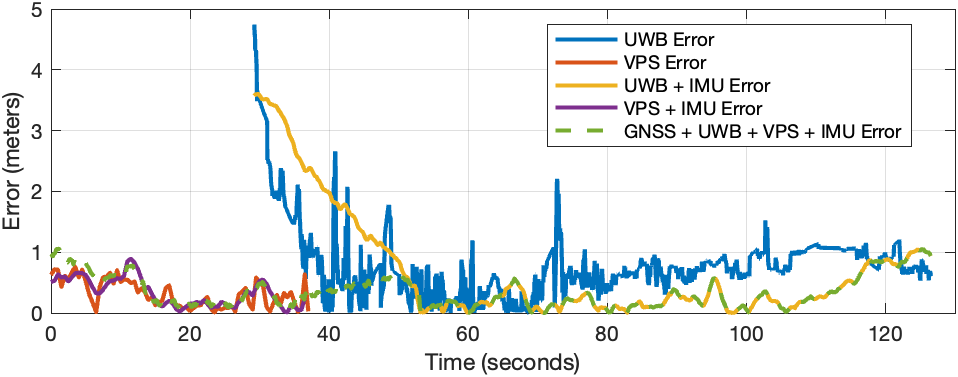}
\caption{Positioning Error Over Time for Various Sensor Fusion Methods.}
\label{fig:time_vs_error}
\end{figure}

Fig.~\ref{fig:time_vs_error} illustrates the dynamic performance of each method. The UWB error fluctuates initially but stabilizes to a lower range. VPS and VPS + IMU methods exhibit consistently low errors, highlighting their robustness. The UWB + IMU combination shows a notable reduction in error compared to standalone UWB. The integrated GNSS + UWB + VPS + IMU approach maintains a consistently low error rate, validating the effectiveness of combining these technologies for optimal positioning accuracy. Note that GNSS and GNSS + IMU data are omitted from Fig.~\ref{fig:time_vs_error} due to their high error values, which would obscure the visualization of other data.

\section{Conclusion}
The study highlights the efficacy of the proposed feasibility study for real-time automated data standardization using large language models, particularly in enhancing seamless positioning. The Intelligent Data Standardization Module (IDSM) achieved near-zero loss and full accuracy in both training and validation phases, demonstrating robust learning capabilities and reliable data standardization across diverse sensor inputs. The Transformation Rules Generation Module (TRGM) significantly reduced the manual effort required for script generation, enhancing productivity and minimizing human intervention. In terms of positioning accuracy, the data fusion approach outperformed traditional standalone methods, achieving a Root Mean Square Error (RMSE) of 0.35 meters and a Mean Absolute Error (MAE) of 0.25 meters. This improvement is critical for applications necessitating precise and reliable location data. The synergy of IDSM's high accuracy and TRGM's automation capabilities presents a powerful solution for sensor data processing, leading to more reliable data and improved decision-making. Despite promising results, the reliance on predefined schemas and manually specified covariance matrices may limit system adaptability in dynamic environments. Additionally, the controlled evaluation setting may not fully capture real-world complexities. Future research should refine this approach for dynamic and complex environmental adaptation and integrate emerging technologies for robustness. Investigate models beyond GPT-4-0613 to understand performance variations, study computational costs and assess real-time versus offline operations.

\section*{Acknowledgment}
This research is supported by the University Grants Committee of Hong Kong under the scheme Research Impact Fund on the project R5009-21 “Reliable Multiagent Collaborative Global Navigation Satellite System Positioning for Intelligent Transportation Systems”.

\vfill


\begin{thebibliography}{20}
\bibliographystyle{IEEEtran}

\bibitem{ref1}
H. Liu, H. Darabi, P. Banerjee and J. Liu, "Survey of Wireless Indoor Positioning Techniques and Systems," in IEEE Transactions on Systems, Man, and Cybernetics, Part C (Applications and Reviews), vol. 37, no. 6, pp. 1067-1080, Nov. 2007, doi: 10.1109/TSMCC.2007.905750.

\bibitem{ref2}
L. Atzori, A. Iera, and G. Morabito, “The Internet of Things: A survey,” Comput. Netw., vol. 54, no. 15, pp. 2787–2805, 2010.

\bibitem{ref3}
J. Hightower and G. Borriello, "Location systems for ubiquitous computing," in Computer, vol. 34, no. 8, pp. 57-66, Aug. 2001, doi: 10.1109/2.940014.

\bibitem{ref4}
C. Luo et al., “Pallas: Self-bootstrapping fine-grained passive indoor localization using WiFi monitors,” IEEE Trans. Mobile Comput., vol. 16, no. 2, pp. 466–481, Feb. 2017.

\bibitem{ref5}
M. Siekkinen, M. Hiienkari, J. K. Nurminen and J. Nieminen, "How low energy is bluetooth low energy? Comparative measurements with ZigBee/802.15. 4", Proc. IEEE Wireless Commun. Netw. Conf. Workshops (WCNCW’12), pp. 232-237, 2012.

\bibitem{ref6}
L. Ojeda and J. Borenstein, "Personal Dead-reckoning System for GPS-denied Environments," 2007 IEEE International Workshop on Safety, Security and Rescue Robotics, Rome, Italy, 2007, pp. 1-6, doi: 10.1109/SSRR.2007.4381271.

\bibitem{ref7}
P. D. Groves, Principles of GNSS, inertial, and multisensor integrated navigation systems, Second edition. Boston: Artech House, 2013.

\bibitem{ref9}
S. Knauth and A. Koukofikis, "Smartphone positioning in large environments by sensor data fusion, particle filter and FCWC," 2016 International Conference on Indoor Positioning and Indoor Navigation (IPIN), Alcala de Henares, Spain, 2016, pp. 1-5, doi: 10.1109/IPIN.2016.7743706.

\bibitem{ref11}
Y. Zhuang et al., “Multi-sensor integrated navigation/positioning systems using data fusion: From analytics-based to learning-based approaches,” Information Fusion, vol. 95,pp. 62–90, Jul. 2023, doi: 10.1016/j.inffus.2023.01.025.

\bibitem{ref14}
Y.Chang et al., “A Survey on Evaluation of Large Language Models,” ACM transactions on intelligent systems and technology, 2024, doi: 10.1145/3641289.

\bibitem{ref16}
“ZED-F9P module,” U-blox. https://www.u-blox.com/en/product/zed-f9p-module (accessed May 21, 2024).

\bibitem{ref17}
“UWB High-Precision Positioning: LinkTrack P-A Series,” Nooploop. https://www.nooploop.com/en/linktrack/ (accessed May 21, 2024).

\bibitem{ref18}
P. -E. Sarlin, C. Cadena, R. Siegwart and M. Dymczyk, "From Coarse to Fine: Robust Hierarchical Localization at Large Scale," 2019 IEEE/CVF Conference on Computer Vision and Pattern Recognition (CVPR), Long Beach, CA, USA, 2019, pp. 12708-12717, doi: 10.1109/CVPR.2019.01300.

\bibitem{ref19}
P. -E. Sarlin, D. DeTone, T. Malisiewicz and A. Rabinovich, "SuperGlue: Learning Feature Matching With Graph Neural Networks," 2020 IEEE/CVF Conference on Computer Vision and Pattern Recognition (CVPR), Seattle, WA, USA, 2020, pp. 4937-4946, doi: 10.1109/CVPR42600.2020.00499.

\bibitem{ref20}
K. T. H. Choi, “tszheichoi / awesome-sensor-logger,” GitHub. https://github.com/tszheichoi/awesome-sensor-logger/.

\end{thebibliography}
\end{document}